\documentclass[review]{elsarticle}

\usepackage{lineno,hyperref}
\usepackage{float}
\usepackage{amsmath,amssymb,amsfonts}
\usepackage{algorithmic}
\usepackage{graphicx}
\usepackage{textcomp}
\usepackage{siunitx}

\usepackage[svgnames,table]{xcolor}
\usepackage{float}
\usepackage{booktabs}
\usepackage{mathtools}
\usepackage{subcaption}
\usepackage{multirow}
\hypersetup{hidelinks}

\usepackage{soul}
\usepackage{caption, copyrightbox}

\modulolinenumbers[5]

\journal{Journal of \LaTeX\ Templates}

\setcitestyle{authoryear,open={(},close={)}}

\begin{document}

\begin{frontmatter}

\title{Learn-Morph-Infer: a new way of solving the inverse problem for brain tumor modeling}


\author[tum,translatum]{Ivan Ezhov\fnref{myfootnote}}
\ead{ivan.ezhov@tum.de}
\fntext[myfootnote]{Corresponding author.}

\author[tum]{Kevin Scibilia}
\author[MechDept]{Katharina Franitza}
\author[tum]{Felix Steinbauer}
\author[tum,translatum]{Suprosanna Shit}
\author[QUZH,translatum]{Lucas Zimmer}
\author[HMS,Broad,Dana]{Jana Lipkova}
\author[tum,translatum,Neuro]{Florian Kofler}
\author[tum,translatum]{Johannes Paetzold}
\author[Mevis]{Luca Canalini}
\author[tum]{Diana Waldmannstetter}
\author[tum,translatum]{Martin J. Menten}
\author[Neuro,translatum]{Marie Metz}
\author[translatum,Neuro]{Benedikt Wiestler\fnref{footnote}}
\author[QUZH]{Bjoern Menze\fnref{footnote}}
\fntext[footnote]{Contributed equally as senior authors.}

\address[tum]{Department of Informatics, TUM, Munich, Germany}
\address[translatum]{TranslaTUM - Central Institute for Translational Cancer Research, TUM, Munich, Germany}
\address[MechDept]{Department of Mechanical Engineering, TUM, Munich, Germany}
\address[QUZH]{Department of Quantitative Biomedicine, UZH, Zurich, Switzerland}
\address[HMS]{Department of Pathology, Brigham and Women's Hospital, Harvard Medical School, Boston, MA}
\address[Broad]{Broad Institute of Harvard and MIT, Cambridge, MA}
\address[Dana]{Data Science Program, Dana-Farber Cancer Institute, Boston, MA}
\address[Mevis]{Fraunhofer MEVIS, Bremen, Germany}
\address[Neuro]{Neuroradiology Department of Klinikum Rechts der Isar, TUM, Munich, Germany}

\begin{abstract}
Current treatment planning of patients diagnosed with a brain tumor, such as glioma, could significantly benefit by accessing the spatial distribution of tumor cell concentration. Existing diagnostic modalities, e.g. magnetic resonance imaging (MRI), contrast sufficiently well areas of high cell density. In gliomas, however, they do not portray areas of low cell concentration, which can often serve as a source for the secondary appearance of the tumor after treatment. To estimate tumor cell densities beyond the visible boundaries of the lesion, numerical simulations of tumor growth could complement imaging information by providing estimates of full spatial distributions of tumor cells. Over recent years a corpus of literature on medical image-based tumor modeling was published. It includes different mathematical formalisms describing the forward tumor growth model. Alongside, various parametric inference schemes were developed to perform an efficient tumor model personalization, i.e. solving the inverse problem. However, the unifying drawback of all existing approaches is the time complexity of the model personalization which prohibits a potential integration of the modeling into clinical settings. In this work, we introduce a deep learning based methodology for inferring the patient-specific spatial distribution of brain tumors from T1Gd and FLAIR MRI medical scans. Coined as \textit{Learn-Morph-Infer}, the method achieves real-time performance in the order of minutes on widely available hardware and the compute time is stable across tumor models of different complexity, such as reaction-diffusion and reaction-advection-diffusion models. We believe the proposed inverse solution approach not only bridges the way for clinical translation of brain tumor personalization but can also be adopted to other scientific and engineering domains. 
\end{abstract}


\end{frontmatter}


\section{Introduction}
\label{sec:introduction}
Glioblastoma (GBM) is the most aggressive brain tumor, characterized by varying and unknown infiltration into the surrounding tissue. After resection of the tumor mass visible in MRI scans, current treatment includes radiotherapy targeting tissue around the visible lesion to account for residual tumor cells. Tumor recurrence is however present in most cases, possibly due to patient-specific and non-uniform distribution of residual tumor cells. Personalization of the clinical (irradiation) target volume could spare more healthy tissue and increase progression-free survival by potentially avoiding recurrence \citep{stupp2014high,harpold2007evolution,Jackson_2015,jana_tmi}.  

Current computational approaches for personalizing radiotherapy planning often rely on solving an inverse problem for GBM growth models \citep{hogea_rd_mass,ender_eikonal,geremia2012brain,bjoern_ipmi,Le_tmi,jana_tmi,scheufele2020automatic,subramanian2020multiatlas,hormuth2021image,lorenzo2021quantitative}. In this context, the growth (forward) models are based on partial differential equations (PDEs) that describe the evolution of tumor cell density in the brain anatomy. The inverse model aims to identify free parameters of the forward model that best match the observation, e.g. tumor outlines from medical imaging modalities. To identify such parameters, the inverse problem can be cast as constrained optimization \citep{hogea_rd_mass,mang2012biophysical,scheufele2019coupling} or Bayesian inference formulations \citep{bjoern_ipmi,jana_tmi,Ezhov2020GeometryawareNS,Ezhov_2019}. 

Predominantly, existing forward GBM models view tumor progression at the macroscopic level by describing gross biomechanical phenomena. These include diffusive motion and proliferation of tumor cells (under simplistic reaction-diffusion PDEs) \citep{bjoern_ipmi}, interaction between the tumor and surrounding tissue (i.e. mass-effect) \citep{subramanian2020multiatlas}, necrotic core formation \citep{patel2017image}, etc. Despite methodological advances in computing the inverse model, the total time for model personalization is still large amounting to multiple hours using such simplistic forward models \citep{subramanian2020multiatlas,hormuth2018mechanically,scheufele2019coupling}. For example, in \citep{scheufele2019coupling} the authors exploit a highly efficient quasi-Newton optimization scheme to infer parameters of the reaction-diffusion model. The inference converges after $\sim$5 hours of compute on 11 dual-x86 CPU nodes for $256^3$ resolution. In \citep{subramanian2020multiatlas}, the mass-effect model is solved using an analogous optimization scheme but implemented on a GPU leading to the same order of compute time for the $256^3$ grid (and up to 1 hour for $128^3$ resolution). Bayesian methods providing uncertainty estimate of the parametric inference \citep{jana_tmi} can take an even longer time (up to days) of computing on specialized CPU clusters. 

Recently, machine learning solutions entered the field of PDEs. Learnable solutions for both forward \citep{raissi2019physics,sitzmann2020implicit,stevens2020finitenet,thuerey2020deep,kasim2020up,Kim_2019} and inverse \citep{papamakarios_snpa,lueckmann_snpb,dax2021real} models were developed. In \citep{papamakarios_snpa,lueckmann_snpb}, the authors proposed a Bayesian framework that allows bypassing a compute of intractable likelihoods necessary for the Bayesian inference. This notably speeds up the inference, however, it still requires thousands of simulations to be generated by the forward model, which in turn can still result in many hours of compute for the inverse problem. In concurrent to our work \citep{dax2021real}, a solution to the inverse problem providing point estimates was proposed. This is achieved via learning a mapping from the physical model simulations directly to the model parameters by leveraging access to large simulated data. Unfortunately, this method is not capable to deal with the varying geometry of the simulation domain. Moreover, as our paper shows, direct mapping to physical model parameters can result in a deterioration of prediction accuracy since the inverse mapping between simulations and parameters is not bijective. 

Few machine learning approaches \citep{Ezhov2020GeometryawareNS,pati2020tmod} also appeared in the brain tumor modeling context. However, \citep{Ezhov2020GeometryawareNS} requires a vast amount of forward model evaluations for convergence of parametric estimation under Bayesian settings for each new patient. In turn, \citep{pati2020tmod} requires access to a dataset of inferred model parameters that can become prohibitively expensive to collect with the growing complexity of the tumor model.

Potential integration of brain tumor modeling into clinical practice would require access to a large cohort of longitudinal clinical data, allowing to estimate the clinical value of a patient outcome's forecast by the tumor models \citep{yankeelov2013clinically}. Integration of current macroscopic models would also require a thorough analysis of forecast consistency between the macroscopic description and higher complexity microscopic models encompassing subcellular biophysics. For this, it is paramount to bring the computational cost of the inverse modeling to a reasonable time. Here, we propose a neural network based methodology for predicting a patient-specific spatial distribution of GBM (from single time-point medical scans, namely T1Gd and FLAIR MRI) that requires neither sampling nor optimization. As such, it is potentially well suited for rapid model personalization in a clinical workflow, but also for scaling up preceding feasibility studies to large patient cohorts. 

Our contribution is a) a pipeline allowing to perform tumor model personalization in a fixed for all patients space, and b) a special learning setup for the network (the main part of the pipeline) performing inverse model inference.  
The method achieves lightning-fast performance in the order of minutes on widely available hardware and the compute time is stable across tumor models of different complexity. This in turn opens the possibility of rapid testing various biophysical models on a large dataset of patients and hence bridging the way for clinical translation. 

\begin{figure*}[h]
\includegraphics[width=1.0\textwidth]{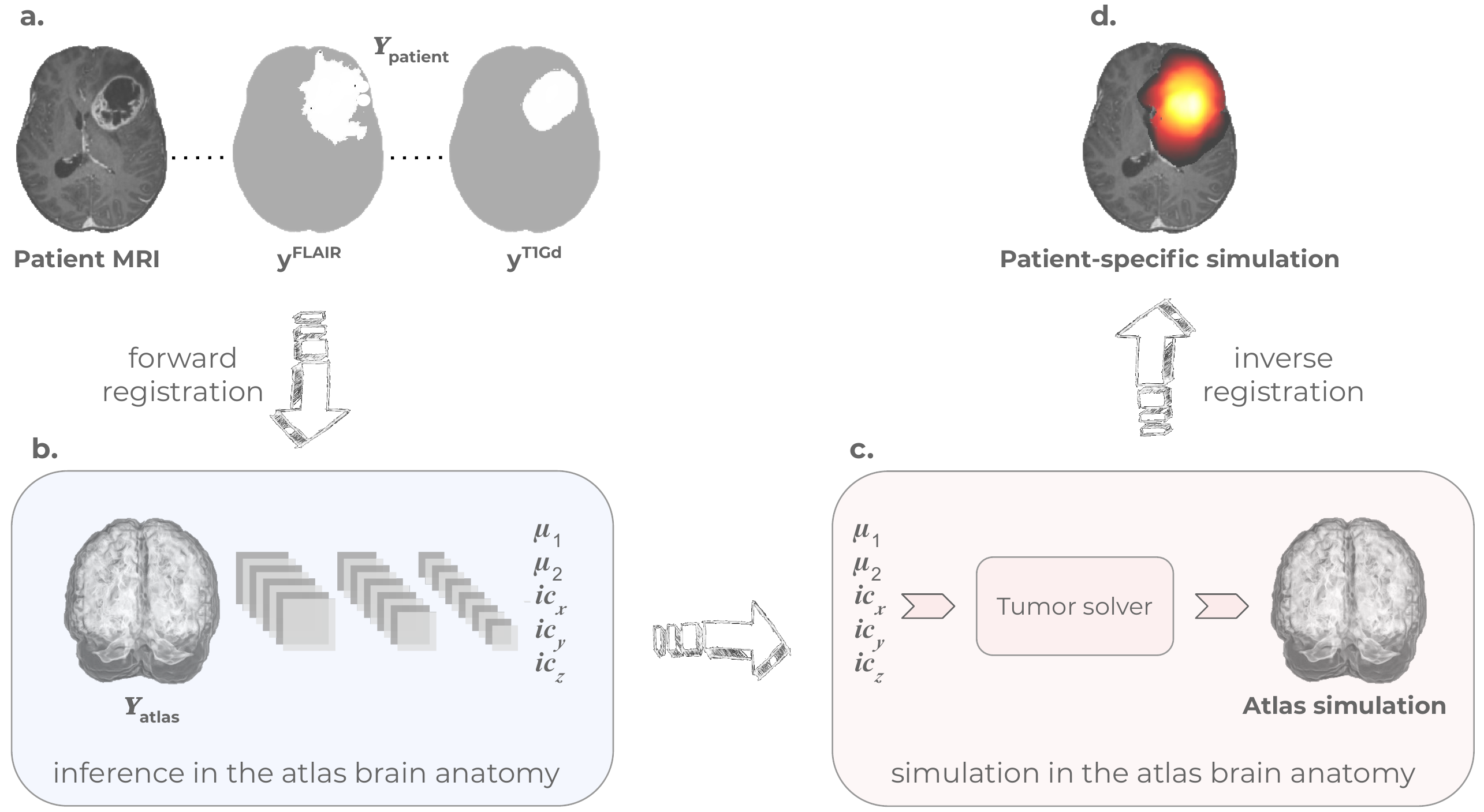}
\caption{\small{A sketch of the inference procedure of the \emph{Learn-Morph-Infer} pipeline. First, the patient's brain tumor segmentations $\{\mathbf{y}^{T1Gd}$, $\mathbf{y}^{FLAIR}\}$ are morphed to the brain atlas space (a). A network trained on synthetic data inputs the morphed segmentations and outputs corresponding tumor model parameters $\{\mu_1, \mu_2, ic_x, ic_y, ic_z\}$ (b). The inferred parameters are used to simulate a tumor in the atlas space (c). Finally, the simulated tumor is morphed back to the patient space (d).}} \label{fig2}
\end{figure*}

\section{Method}
\label{sec:method}

\subsubsection{Learn-Morph-Infer Pipeline}
 
In contrast to a few works like \citep{hormuth2019calibrating,Ezhov_2019,tuncc2021modeling}, we solve the inverse problem by relying only on single time-point MRI observations of the brain tumor $Y = \{ \mathbf{y}^{T1Gd}, \mathbf{y}^{FLAIR}\}$. This is the most realistic clinical scenario as normally treatment follows immediately after the first MRI scanning. Given a tumor observation $Y$, our goal is to calibrate a set of personalized parameters $\theta_c$ of the forward tumor growth model that infers the underlying patient-specific tumor cell density $c_{patient}$ in the patient anatomy\footnote{Let us explain how we see the potential clinical utility of glioma modeling. In clinical practice, for glioma patients normally only a single MRI scan is taken before treatment. It is not feasible to capture the dynamics of the pathology by calibrating a tumor model from a single scan - multiple sets of model parameters can result in the same simulated tumor profile. Thus, one cannot reliably predict tumor evolution over time as different sets of parameters will result in tumor trajectories diverging over time.
Instead, we are only interested in identifying a simulated tumor whose morphology (and, hence, tumor cell distribution) best matches the morphology of the tumor we see on a single MRI scan. The best matching simulated profile will then serve to predict tumor infiltration beyond contrast-enhancing tumor areas and thereby define the target volume for radiotherapy (as opposed to the currently used simple EORTC criteria - 2cm border surrounding the cavity \citep{young1999measurement}).
}. For a given patient, this is achieved via the proposed \emph{Learn-Morph-Infer} pipeline:

\begin{itemize}
\item We register a patient MRI image to the brain atlas  \citep{rohlfing2010sri24} and obtain a transformation matrix.

\item The transformation matrix is used to morph scans based on the patient anatomy $Y_{patient}$ to scans in the atlas anatomy $Y_{atlas}$ as illustrated in Fig. \ref{fig2}a-b.

\item A neural network, that has learned to solve the inverse problem $Y_{atlas} \rightarrow \theta_c$ through prior training on simulated data $Y_{sim}$ (Fig. 1), predicts $\theta_c$ during inference, Fig. \ref{fig2}b.

\item These parameters $\theta_c$ are used as input in the tumor growth model (forward solver) to infer a tumor cell density $c_{atlas}$ in the atlas space, Fig. \ref{fig2}c.

\item The tumor volume $c_{atlas}$ is transformed back to the patient space with the inverse transformation matrix, yielding $c_{patient}$ as displayed in Fig. \ref{fig2}d.
\end{itemize}

\subsubsection{Forward tumor model}
We independently probe two types of non-linear PDEs describing tumor growth:  

\paragraph{Reaction-diffusion equation} First, we consider the Fisher-Kolmogorov PDE describing the evolution of the tumor cell concentration $c$ by considering cell diffusion and proliferation,

\begin{equation}
\frac{\partial c}{\partial t} = \nabla\cdot(\mathbf{D} \nabla c) + \rho c(1-c)  
\label{eqn:reac-diff}
\end{equation}

\begin{equation}
\nabla c \cdot \textbf{n} = 0 \quad boundary\ condition
\end{equation}

\noindent Here, $\rho$ denotes the tumor proliferation rate while the infiltrative behaviour of the tumor is modelled by the diffusion tensor $\mathbf{D} = D\cdot\mathbb{I}$. The equation is solved in a three dimensional atlas brain anatomy segmented into white matter (WM), grey matter (GM) and cerebrospinal fluid (CSF). The diffusion coefficient $D$ is computed for each voxel $\mathbf{i}$ with location $(i_x,i_y,i_z)$  as $D_i = p_{w_i} D_w +p_{g_i} D_g$, where $p_{w_i}$, $p_{g_i}$ describe percentages at voxel $\mathbf{i}$ and $D_w$, $D_g$ diffusion coefficients of WM and GM respectively, and a relation $D_w = 10\cdot D_g$ is assumed \citep{jana_tmi}. No cell diffusion into CSF is feasible according to the model. The solver based on this growth model takes  $\theta_c = \{D_w, \rho, T, ic_x, ic_y, ic_z\}$ as input and returns a tumor cell density $c$. The parameters $\mathbf{x}=( ic_x, ic_y, ic_z)$ define the initial condition where the tumor is initialized at time $t$=0 as a point seed. The tumor is simulated until the time of detection $T$.

\paragraph{Reaction-diffusion-advection equation} The second type is a non-linear reaction-diffusion-advection PDE analogous to \citep{subramanian2020multiatlas}. In the following, the brain tissue will be represented by $\mathbf{m} = (m_{WM}(\mathbf{i},t),\ m_{GM}(\mathbf{i},t),\ m_{CSF}(\mathbf{i},t))$ for each voxel $\mathbf{i}$ and time $t$. The normalized tumor cell density $c=c(\mathbf{i},t)$  can be modelled by the following equations:

\begin{equation}
\frac{\partial c}{\partial t} = \nabla\cdot(\mathbf{D} \nabla c) - \nabla(c\mathbf{v}) + \rho c(1-c)
\label{eqn:reac-diff-adv}
\end{equation}
\begin{equation} 
\frac{\partial \mathbf{m}}{\partial t} + \nabla\cdot(\mathbf{m} \otimes \mathbf{v} )  = 0 
\end{equation}
\begin{equation}
\nabla\cdot (\lambda  \nabla \mathbf{u}  + \mu (\nabla \mathbf{u} + \nabla \mathbf{u}^\top )) = \gamma  \nabla c 
\label{eqn:displacement}
\end{equation}
\begin{equation}
\frac{\partial \mathbf{u}}{\partial t} = \mathbf{v}
\label{eqn:dudt}
\end{equation}
\begin{equation}
\nabla c \cdot \textbf{n} = 0 \quad boundary\ condition
\end{equation}
\begin{equation}
\mathbf{m} = 0 \quad boundary\ condition
\end{equation}
\begin{equation}
\mathbf{u} = 0 \quad boundary\ condition
\end{equation}
\begin{equation}
\mathbf{v} = 0 \quad boundary\ condition
\end{equation}

\noindent Coupling Eqn. \ref{eqn:reac-diff-adv} to a linear elasticity model, Eqn. \ref{eqn:displacement}, allows considering deformation in the anatomy due to a mass effect induced by tumor growth \citep{subramanian2019simulation}. The linear elasticity model is defined by the Lamè coefficients $\lambda$ and $\mu$ as specified in Eqn. \ref{eqn:displacement}. The displacement $\mathbf{u}$ is represented in the advection term of Eqn. \ref{eqn:reac-diff-adv}. The degree of the mass effect depends on the selection of the mass effect parameter $\gamma$. 

\begingroup
\setlength{\tabcolsep}{10pt} 
\renewcommand{\arraystretch}{0.7} 
\begin{table*}
\centering
\begin{tabular}{*{3}c}
\rowcolor{LightBlue!40}
Parameter symbol & \vline & Parameter meaning   \\ \hline
$c$ & \vline & Tumor cell density  \\\hline
$\mathbf{D}$ & \vline & Diffusion tensor \\\hline
D & \vline & Diffusion coefficient  \\\hline
$\mathbb{I}$ & \vline & Identity tensor  \\\hline
$\rho$ & \vline & Proliferation rate  \\\hline
$\textbf{n}$ & \vline & Unit vector normal to boundary  \\\hline
$\mathbf{u}$ & \vline & Advection displacement  \\\hline
$\mathbf{v}$ & \vline & Advection velocity  \\\hline
$\mathbf{m}$ & \vline & Brain tissue maps  \\\hline
$\lambda, \mu$ & \vline & Lamè coefficients  \\\hline
$\gamma$ & \vline & Mass effect parameter  \\\hline
\end{tabular}
\caption{Description of parameters driving the reaction-diffusion and reaction-advection-diffusion PDEs.}
\end{table*}
\endgroup

\subsubsection{Linking cell density with MRI signal}
MRI modalities capture structural information about the brain tumor. T1Gd contrasts the tumor core, whereas FLAIR informs about the area of the edema in addition to the tumor core. It is established practice \citep{matthieu_bayesian,jana_tmi,subramanian2020multiatlas,tunc2021modeling,KONUKOGLU2010111,bjoern_ipmi} to consider binary segmentations corresponding to the MRI scans to inform biophysical models. The binary masks contain zeros in the area of healthy tissues and are non-zero in the tumor-related area. In order to relate the segmentations $Y=\{ \mathbf{y}^{T1Gd}, \mathbf{y}^{FLAIR}\}$ to a simulated tumor cell density $c$, we threshold the density at randomly sampled levels $c_t^{T1Gd}$ and $c_t^{FLAIR}$ ($c_t^{T1Gd} > c_t^{FLAIR}$) to obtain $Y_{sim}=\{ \mathbf{y}^{T1Gd}_{sim}, \mathbf{y}^{FLAIR}_{sim} \}$ reproducing the real segmentations \footnote{We want to clarify here the thresholding step in more detail. The output of our tumor solver is a continuous tumor cell density distribution. The experimental observation to which we want to fit our tumor model simulations comes in a form of binary tumor segmentation. In order to identify a simulation best fitting the binary segmentation, we threshold our simulated continuous tumor cell distribution to obtain a binary volume, which in turn can already be nicely compared with the real patient segmentation. Oftentimes, in literature people resort to fixed values of cell density for the threshold levels. However, there is no understanding (neither agreement) in the community of what levels would best correspond to a real scenario. Different works used different numbers in their studies \citep{Le_tmi}. To consider the most general case, we randomly sampled the values within the ranges reported in the literature.
}.

\subsubsection{Learning the inverse model in atlas space}
\label{learning}

As discussed in the previous section, the key step of the \emph{Learn-Morph-Infer} pipeline is to learn the inverse tumor model using a neural network that can infer a set of personalized parameters $\theta_c$ from corresponding tumor observations $Y_{atlas}$. 
In order to create a dataset for the network training, we generate 100,000 tumors in the atlas space by randomly sampling tumor model parameters $ \{D_w, \rho, T, ic_x, ic_y, ic_z, c_t^{T1Gd}, c_t^{FLAIR}\}$ within physiologically plausible ranges \citep{swanson2000quantitative}. To form the neural network input, the simulated MRI segmentations are combined into one volume $\mathbf{y}^{MRI}_{sim} = 0.666\cdot\mathbf{y}^{T1Gd}_{sim} + 0.333\cdot\mathbf{y}^{FLAIR}_{sim}$.

\textit{Reformulation of the inverse problem.} Now, the question is what should be used as a network prediction? It is tempting to try to predict the tumor model parameters directly. However, it is well known that the inverse problem is highly ill-posed, i.e. numerous sets of dynamic parameters $\{D_w, \rho, T\}$ correspond to the same simulated cell density profile. Thus, we have two sources of prediction error: 
\noindent a) coming from the fact that we learn a mapping from one to many. Imagine we have two sets of dynamic parameters $\{D_w, \rho, T\}$ and $\{D_w^*, \rho^*, T^*\}$ that result in the same tumor profile $\mathbf{y}^{MRI}_{sim}$. If we train a network in a supervised fashion, every time the network predicts $\{D_w^*, \rho^*, T^*\}$ for a $\{D_w, \rho, T\}$-$\mathbf{y}^{MRI}_{sim}$ pair (and vice-versa), it will be falsely penalized; 
\noindent b) the actual error that comes from limited network capacity to accurately learn the mapping.   

\begin{figure}[H]
\centering
\includegraphics[width=1.0\textwidth]{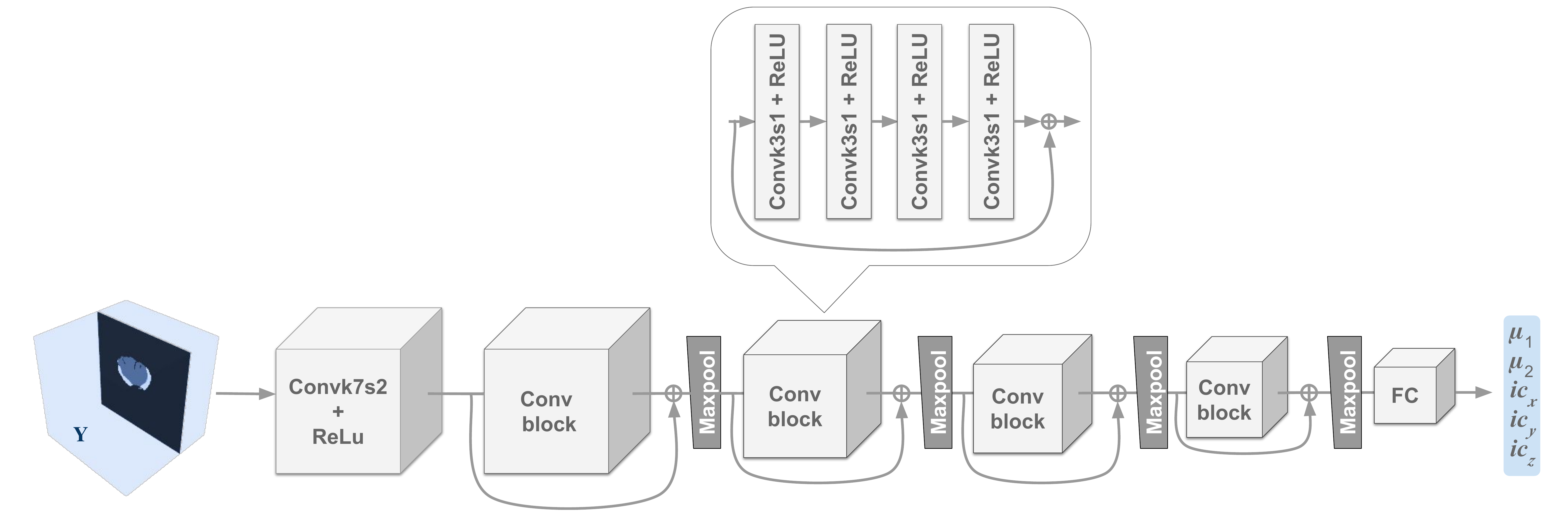}
\caption{\small{A design of the inverse model network - one of the key elements of the \emph{Learn-Morph-Infer} pipeline. The network design represents a ResNet type architecture. It takes as an input the binary brain tumor segmentations $\{\mathbf{y}^{T1Gd}$, $\mathbf{y}^{FLAIR}\}$ and outputs $\{\mu_1, \mu_2, ic_x, ic_y, ic_z\}$}. Crucially, the network predicts $\{\mu_1, \mu_2\}$ which are not the exact forward model parameters but time-independent combinations thereof.} 
\label{fig:net}
\end{figure}

Clearly, the error of type (a) should negatively affect the learning performance, as the network may be penalized for making a sensible prediction. To circumvent this, we do not predict the parameters directly. As evident from Eqn. \ref{eqn:reac-diff}, normalization of the time parameter $T$ in Eqn. \ref{eqn:reac-diff} is equivalent to re-scaling of the proliferation $\rho$ and diffusion $D$ coefficients \citep{subramanian2020did}. This means that for sets $\{D_w, \rho, T\}$ and $\{D_w^*, \rho^*, T^*\}$ corresponding to the same simulated tumor, combinations of time-independent parameters $\mu_1=\sqrt{D_wT}$, $\mu_2=\sqrt{T\rho}$ stay constant 
($\sqrt{D_wT}$=$\sqrt{D_w^*T^*}$, $\sqrt{T\rho}$=$\sqrt{T^*\rho^*}$) \citep{ender_eikonal}. Hence, predicting these combinations of time-independent parameters $\mu_1$ and $\mu_2$ relaxes the error type (a). In order to calculate back the $\{D_w, \rho, T\}$, we introduce a third combination as $v = 2\sqrt{D_w\rho}$. As it is not possible to infer the velocity $v$ from a single time-point observation, we set the velocity equal to the mean velocity of the used sampling range, 200 $mm/year$ (note also that for our purpose the choice of $v$ is irrelevant as any tumor simulation can be obtained with arbitrary $v$ \citep{bjoern_ipmi}). Given $\{\mu_1, \mu_2, v \}$, we can calculate $\{D_w=\frac{\mu_1v}{2\mu_2}, \rho=\frac{\mu_2v}{2\mu_1}, T=\frac{2\mu_2\mu_1}{v}\}$.

\noindent Following this reasoning, we make the network to predict five parameters: $\{\mu_1, \mu_2\}$ and $\{ic_x, ic_y, ic_z\}$. Note that we do not predict $ \{c_t^{T1Gd}, c_t^{FLAIR}\}$, as we do not need the threshold parameters at further steps of the \emph{Learn-Morph-Infer} pipeline.

\textit{The inverse network design.} The network we chose to learn the inverse model represents a convolutional architecture, depicted in Fig. \ref{fig:net}. Every convolution in the network is followed by a rectified linear unit (ReLU) nonlinearity. The input is passed through an initial convolution of kernel size $7$, stride $2$ and $64$ filters (k7s2f64), downsampling the volume to $64^3$ and increasing the number of channels to $64$. This volume of size $64^3$ x $64$ is input through four convolutional blocks, where every convolutional block contains four convolutions of kernel size $3$, stride $1$, and $64$ filters (i.e. number of filters is kept constant throughout the network). A convolutional block uses a skip connection to learn a residual mapping  \citep{deepreslearning}, with the input being added element-wise to the output of the four convolutions. The first three convolutional blocks are followed by a MaxPool3D layer (with parameters k2s2) to downsample the 3D volumes by two. The last convolutional block is followed by a global average pooling layer, shrinking the $64$ $3$D volumes to $64$ neurons that can be linked through a fully connected (FC) layer to the output. These outputs are linearly interpolated into the [-1, 1] range for training. 


\section{Results}
\label{sec:results}

\subsection{Data and implementation details}
\textit{Synthetic data}. The simulated tumors used for training the network have resolution of $128^3$. The simulations were generated by randomly sampling patient-specific parameters from the following ranges:
$D_w \in [\SI[per-mode = fraction]{0.0002}{\cm\squared\per\day}, \SI[per-mode = fraction]{0.015}{\cm\squared\per\day}]$, $\rho \in [\SI[per-mode = fraction]{0.002}{\per\day}, \SI[per-mode = fraction]{0.2}{\per\day}]$, $T \in [50\si{\day},1500\si{\day}]$, $x \in [0.15,0.7]$, $y \in [0.2,0.8]$, $z \in [0.15,0.7]$, $c^{T1Gd}_t \in [0.5, 0.85]$, $c^{FLAIR}_t \in [0.05, 0.5].$  The elasticity model parameters $\lambda, \mu, \gamma$ were taken from \citep{subramanian2019simulation}. Tumors that are unrealistically small or large have been discarded (based on minimum and maximum tumor sizes of real tumors from BraTS dataset \citep{menzebrats}). The simulations dataset is then divided into a training set (80000 tumors), validation set (8000 tumors) and test set (12000 tumors).

\begin{figure}[h!]
\centering
\includegraphics[width=1.0\textwidth]{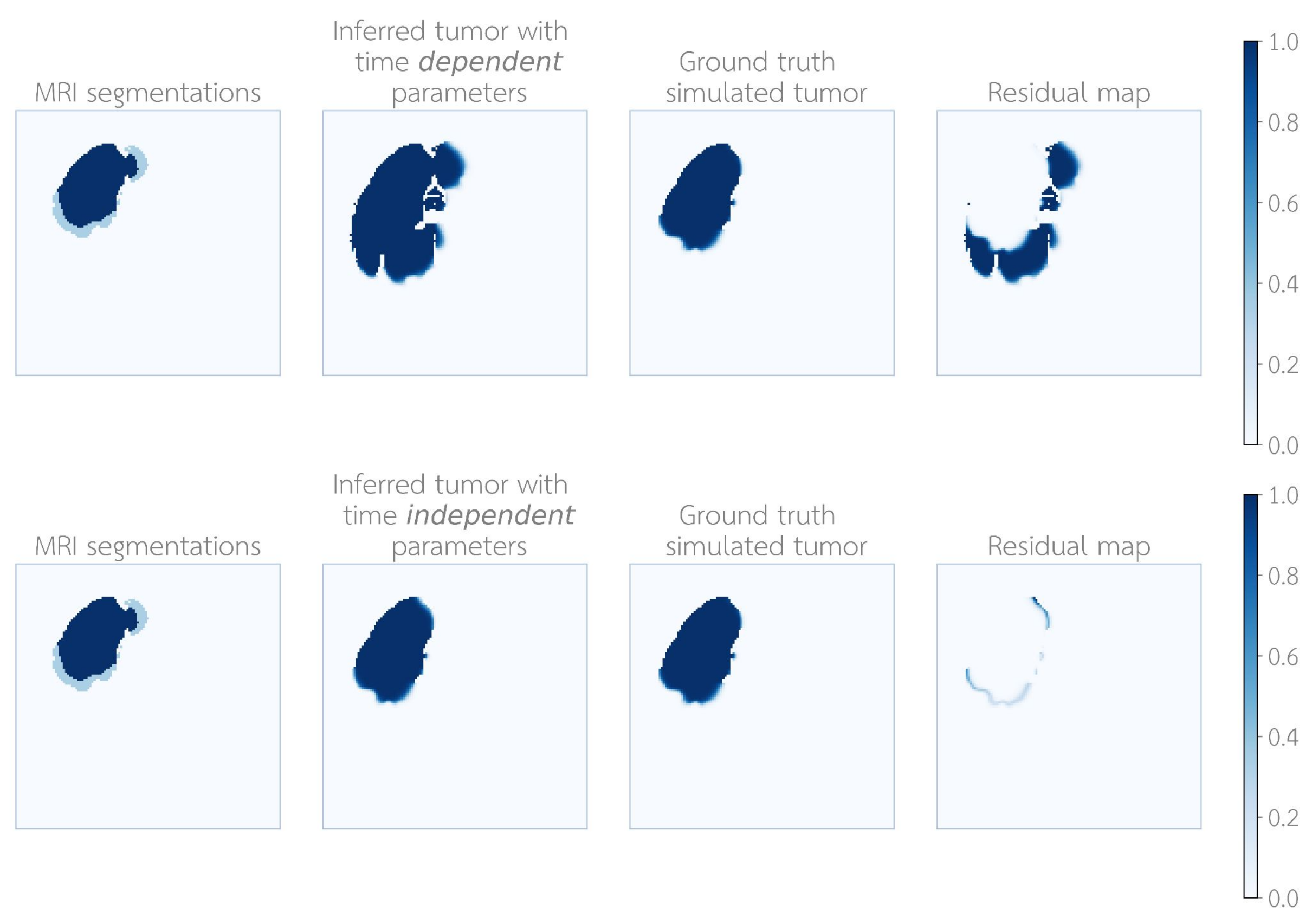}
\caption{\small{Qualitative comparison between the tumors inferred using the \emph{Learn-Morph-Infer} pipeline with two different network training strategies: with time-dependent ($D_w, \rho, T, \mathbf{x}$) parameters as network output, and time-independent ($\mu_1, \mu_2,\mathbf{x}$) parameters as output. The examples correspond to the Fisher-Kolmogorov tumor model with MAE equal to 0.495 (within WM) and 0.496 (within GM) for time-dependent inference, and 0.048 (within WM) and 0.048 (within GM) for time-independent inference. The first column "MRI segmentations" corresponds to the FLAIR+T1Gd segmentation, the second one "Inferred tumor" corresponds to the simulation inferred by the proposed \emph{Learn-Morph-Infer} pipeline, the third column "Ground truth simulated tumor" corresponds to the simulated tumor which we used for the sensitivity analysis (see the "Synthetic data" section for details how these test data were formed), and the fourth "Residual map" column is the difference between the 2nd and 3rd columns' images.}}
\label{qualcomp}
\end{figure}

Analogously to how we created the training data, we form synthetic test data by thresholding a simulated tumor $c_{gt}$ at two levels. Then we pass the obtained thresholded volumes through the pipeline to infer the tumor cell distribution $c_{pred}$. Finally, we quantitatively compare the difference between $c_{gt}$ and $c_{pred}$. 

\textit{Real data}. The data set of real MRI images consists of an in-house cohort of 80 patients with a newly diagnosed glioblastoma, IDH wild type as per the 2016 WHO classification of brain tumors. All patients gave written informed consent to be part of an observational cohort. Preoperative MR images were converted to NIfTI format. For image preprocessing (co-registration, skull stripping) and automated tumor segmentation, we used BraTS Toolkit \citep{kofler2020brats}, a tool we developed locally and which is freely available. The average age of the patient cohort is 60 years with a minimum of 26 and a maximum of 79 years. The cohort is equally represented by gender. 

\textit{Implementation details}.
For the registration between the patient and atlas brain MRI scans, we use the Advanced Normalization Tools (ANTs) \citep{avants2009advanced}. We choose a deformable SyN registration that ensures providing both forward and inverse transformation  with step-size 0.25, weight 1, and region radius for cross-correlation computation r=4. Cross-correlation is used as a similarity metric. The optimization is performed over two resolutions with a maximum of 50 iterations at the coarsest level, and 20 at the final level. The tumor area on the patient scan was masked for the registration. We use a Gaussian regularizer with a sigma of 3 operating on the similarity gradient. These settings provided high morphing quality at relatively fast compute ($\sim$2 minutes).

The network is initialized with He initialization as in the original ResNet architecture \citep{deepreslearning}, and is trained with the AdamW \citep{loshchilov2018decoupled} optimizer, which is a variant of the Adam optimizer \citep{adam} using decoupled weight decay. We use an initial learning rate of $6$ x $10^{-5}$ which is decayed exponentially after every batch by a factor of $0.999997$. Weight decay with a factor of $0.05$ is used as a regularization technique. Furthermore, we train the network with a batch size of $32$ and the Mean Squared Error (MSE) loss function. All training and testing runs were executed on an NVIDIA Quadro RTX 6000 with the PyTorch framework. 

\subsection{Experiments}
We perform two sets of experiments: a) on synthetic data to estimate the accuracy of the learnable inverse model, and b) on patient MRI scans to qualitatively probe the transferability of the method to real data. 

\begin{table}[h]
\begin{center}
 \begin{tabular}{p{0.3cm}  p{0.8cm}  p{2.3cm} | p{0.8cm} p{0.8cm} p{0.8cm}| p{0.8cm} p{0.8cm} p{0.8cm}} 
 \midrule
  \multicolumn{3}{c}{Experiment} &%
  \multicolumn{3}{c}{MAE}&%
  \multicolumn{3}{c}{DICE}\\  
  \midrule
    ID & \shortstack{Tumor \\ Model} & \shortstack{Predicted \\ Parameters} &%
    
    GM &%
    WM &%
    CSF &%
    $c_t$=0.01 &%
    $c_t$=0.1 &%
    $c_t$=0.8 \cr
 \midrule
 1 & \multirow{3}{*}{$FK$} & $\{D_w, \rho, T\},\{\mathbf{x}\}$ & 0.461 & 0.463 & 0.0 & 0.607 & 0.558 & 0.423\\
 2  &  & $\{\mu_1, \mu_2\},\{ \mathbf{x} \}$ & 0.059 & 0.059 & 0.0 & 0.940 & 0.928 & 0.855 \\
 3  &  & $\{\mu_1, \mu_2,\mathbf{x}\}$  & \textbf{0.057} & \textbf{0.057} & \textbf{0.0} & \textbf{0.943} & \textbf{0.932} & \textbf{0.861} \\ 
 \midrule
 4 & \multirow{3}{*}{$ME$} & $\{D_w, \rho, T\},\{\mathbf{x}\}$ & 0.089 & 0.087  & 0.059 & 0.846 & 0.807 & 0.737\\
 5  &  & $\{\mu_1, \mu_2\},\{ \mathbf{x} \}$  & 0.059 & 0.057 & 0.054& 0.877 & 0.841 & 0.772\\ 
 6  &  & $\{\mu_1, \mu_2,\mathbf{x}\}$  & \textbf{0.055} & \textbf{0.054} & \textbf{0.054}& \textbf{0.886} & \textbf{0.850}  & \textbf{0.783}\\ 
\end{tabular}
\caption{Ablation analysis on the test set (12k samples). In total, we perform 6 experiments. First three experiments are performed for the Fisher-Kolmogorov (FK) tumor model: 1) Two separate neural networks predicting $\{ D_w, \rho, T\}$ and $\{\mathbf{x}=( ic_x, ic_y, ic_z)\}$ , 2) Two separate neural networks for prediction of the growth $\{ \mu_1,\mu_2\}$ and location $\{\mathbf{x}\}$  parameters, 3) Single neural network predicting $\{\mu_1,\mu_2, \mathbf{x}\}$. The last experiments 4-6 are analogous but performed for the mass-effect (ME) model. 
As error measure for all experiments, we use the mean absolute error (MAE) $||c_{sim}-c_{gt} ||$  in WM, GM, and CSF, as well as DICE score between ground truth and simulated tumor cell density thresholded at different $c_t$. The non-zero error in the CSF area for the ME model comes from the fact that the simulated tumor is allowed to displace the healthy tissue including the CSF. For both FK and ME models, the usage of a single network predicting time-independent parameters results in a notable increase in accuracy compared to the time-dependent counterpart.}
\end{center}
\label{experiments}
\end{table}

\subsection{Synthetic test} 
In Tab. 2, for two different tumor models, we show results of the ablation analysis, wherein we perform multiple experiments varying the neural network input and output configurations. First, we provide empirical proof that a network predicting $\{D_w, \rho, T\}$ instead of time-independent parameter combinations $\{\mu_1, \mu_2\}$ cannot be trained reliably. Mean absolute error, as well as DICE score, improve significantly when the network predicts $\{\mu_1, \mu_2\}$ (for the ME model the improvement is less pronounced that we attribute to a higher numerical error in our forward solver implementation for this model). Second, we tested whether the performance can be improved by learning two separate networks predicting growth $\{ \mu_1, \mu_2 \}$ and initial location parameters $\{\mathbf{x}=( ic_x, ic_y, ic_z)\}$, respectively. This test did not reveal an improvement compared to a single network predicting all parameters.

\begin{figure*}[h!]
\centering
\includegraphics[width=0.8\textwidth]{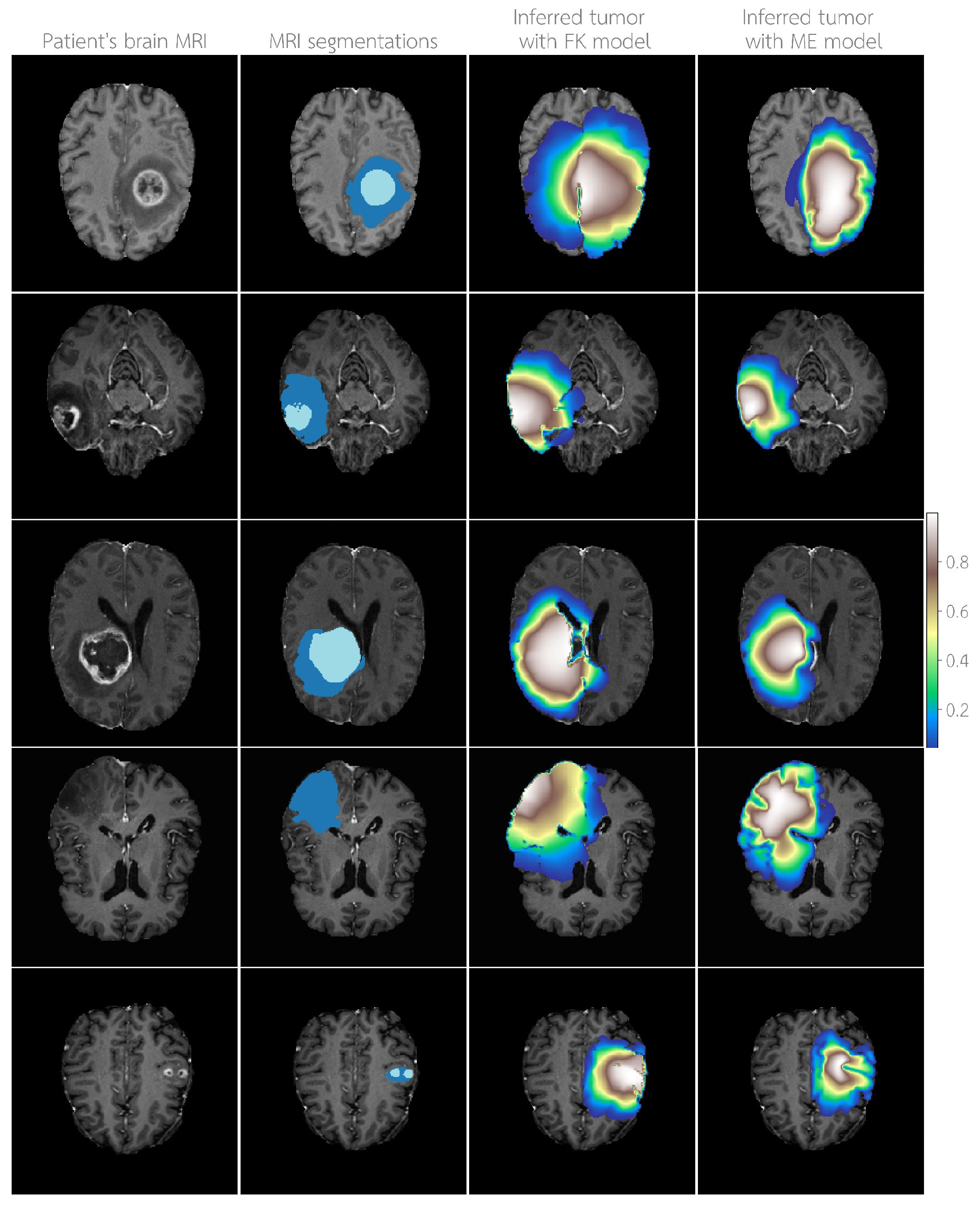}
\caption{\small{Examples of patient-specific simulations produced using the \emph{Learn-Morph-Infer} method. The inverse model network was trained on samples from the Fisher-Kolmogorov (3rd column) and mass effect (4th column) forward models.}} \label{realinf}
\end{figure*}

Fig. \ref{qualcomp} qualitatively showcases the accuracy of inference using the \emph{Learn-Morph-Infer} pipeline. As discussed before, depending on the network training strategy (either predicting time-dependent $\{D_w, \rho, T, \mathbf{x}\}$ parameters, or time-independent $\{\mu_1, \mu_2,\mathbf{x}\}$ parameters as network output), the accuracy of the final simulated tumor notably differs. If the proposed learning with time-independent combinations of parameters provides close to the ground truth tumor profile, then the learning with time-dependent combinations makes the inference hardly useful.

\begin{figure*}[h!]
\centering
\includegraphics[width=1.0\textwidth]{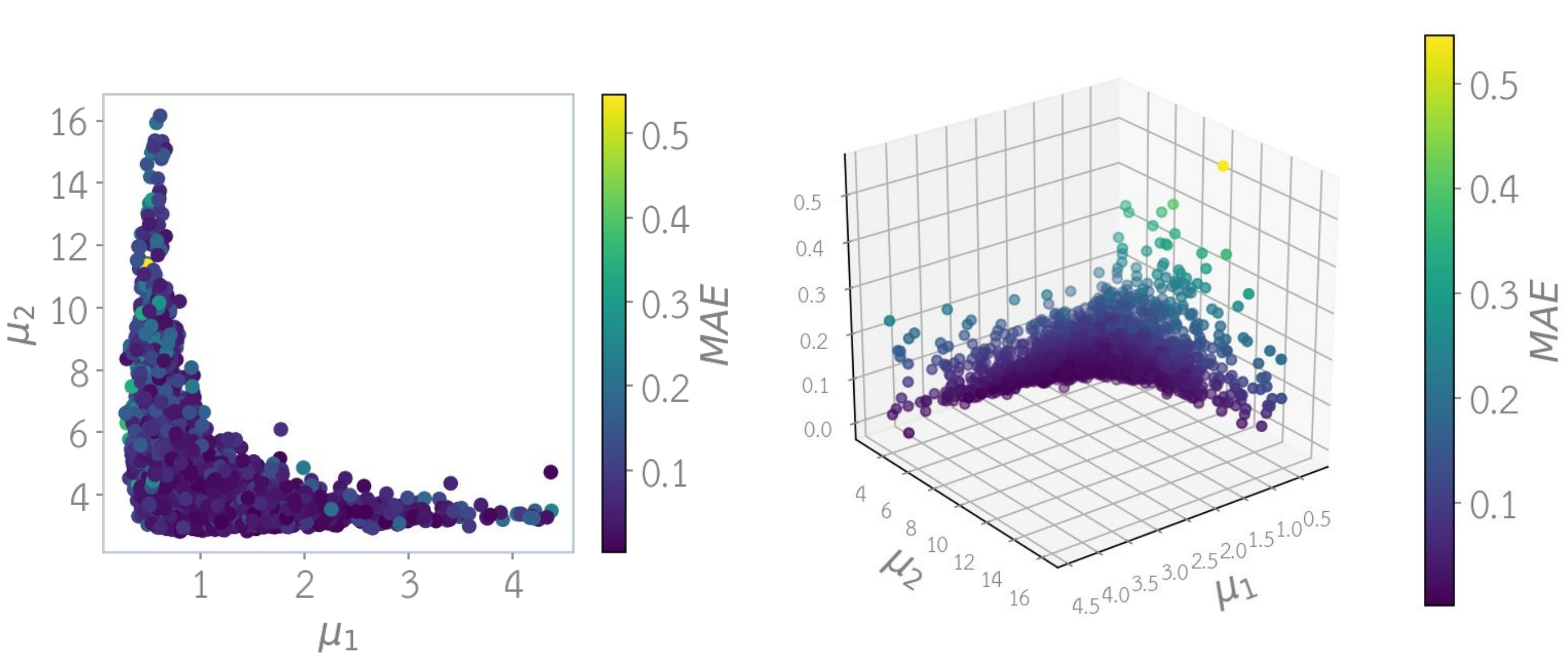}
\caption{\small{Distribution of the mean absolute error over the range of values for the parameters $\mu_1, \mu_2$, in 2D (left) and 3D (right) views.}} \label{errdist}
\end{figure*}

Finally, we also analyzed the robustness of the proposed inverse network against the wide range of model parameters. Fig. \ref{errdist} demonstrates the distribution of the mean absolute error over the range of values for the parameters $\mu_1, \mu_2$ pertaining to the test set of the FK model. From these scatter plots, we conclude that the network performance is stable across the ranges used for the test set.

\subsection{Real MRI patient data}
We performed qualitative validation of the method on a large cohort of brain tumor patients who underwent MRI testing. Binary segmentation corresponding to the T1Gd and FLAIR modalities were used as input to the \emph{Learn-Morph-Infer} method. Fig. \ref{realinf} showcases examples of inferred tumor simulations for various tumor grades and locations in patients' brains. Out of 80 cases, there were less than 10\% cases in which the pipeline outputted wrong results: tumor occupying the whole brain or absence of tumor.

\begin{figure*}[b!]
\centering
\includegraphics[width=1.0\textwidth]{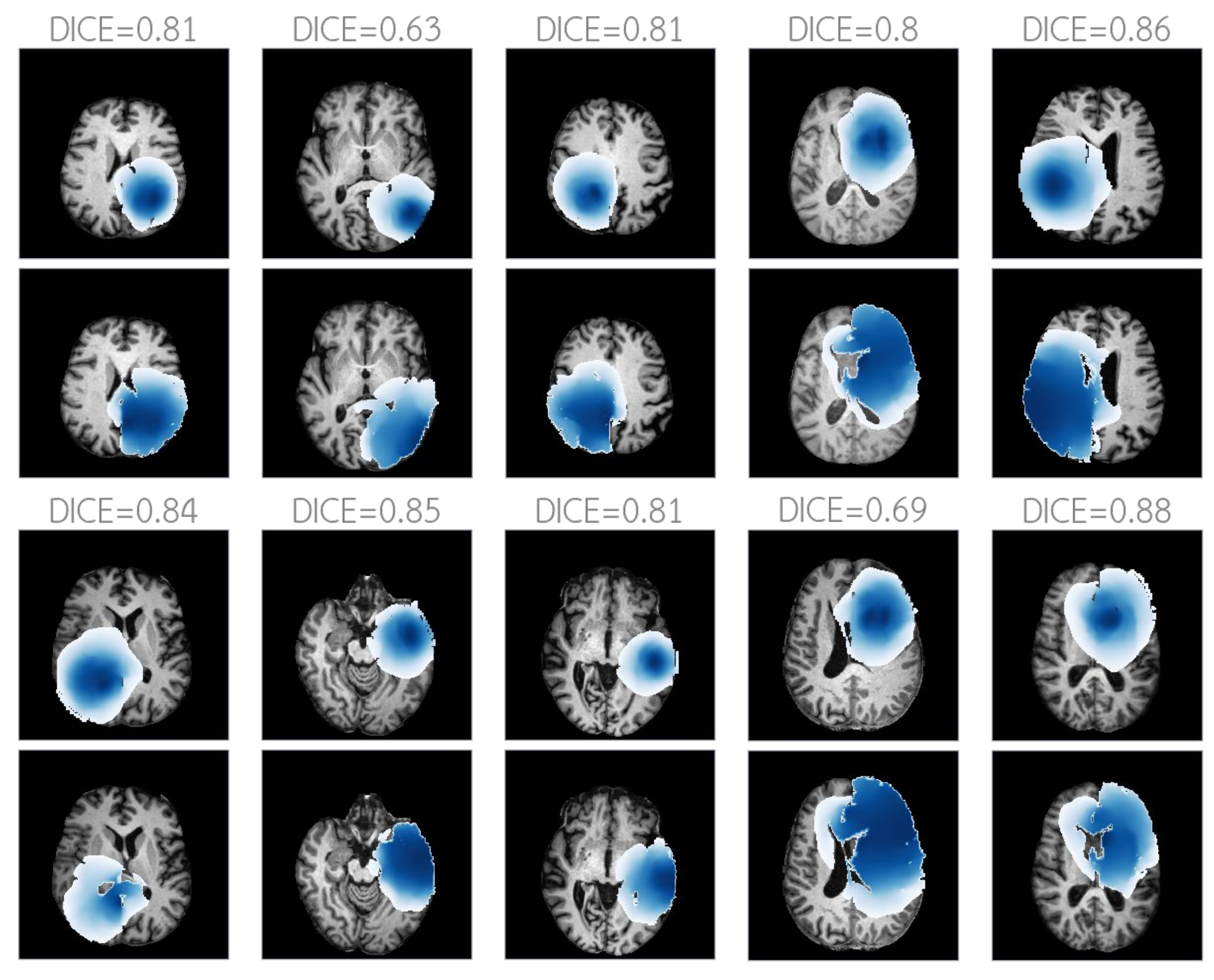}
\caption{\small{Examples of patient-specific simulations produced using the \emph{Learn-Morph-Infer} method (above) and the MCMC-based glioma solver from \citep{jana_tmi} (below) for the FK model. Despite of the different nature of the methods and inevitable errors coming from the network misprediction, the DICE score for most of the cases is around 0.8 (the DICE was computed after thresholding the tumor cell concentration at $c_t=0.01$).}} \label{realinf2}
\end{figure*}

To quantitatively estimate the accuracy of the proposed method, we compared it with the publicly available glioma solver from \citep{jana_tmi}. Due to the computationally costly inference of the latter, we evaluated the performance on 10 randomly chosen glioma patients. The results of the comparison are provided in Fig. \ref{realinf2}. For most of the cases, the overlap in terms of DICE is around 0.8. Note that this error includes not only the contribution from network misprediction and registration back and forth to the brain atlas space but likely also from just different nature of the methods (our proposed method provides point estimates, while the glioma inverse solver by \citep{jana_tmi} is a probabilistic method). A natural extension of our work to reduce the error would be to run the sampling-based (or optimization) method in patient space after performing inference with the proposed \textit{Learn-Morph-Infer} method but within narrow parametric ranges centered around estimates predicted by our learnable method. Or even a simpler extension - instead of the steps "c" and "d" in Fig.1, one can run the tumor solver directly in patient space with model parameters inferred at step "b" (such procedure would mitigate the error coming from the mismatch between brain atlas and patient anatomy). 

\subsection{Computing time} 
The total time including registration, morphing to atlas space, inference, tumor simulation, and morphing back to patient space is 4-7 minutes. The time for the inverse model network's inference is around 2 seconds for $128^3$ resolution. Crucially, the inference time for the more complicated model with mass effect stays the same as for the Fisher-Kolmogorov tumor model. This emphasizes the key practical contribution of the proposed method in that it allows constant time model personalization for an arbitrary tumor model complexity.

\section{Conclusion}
\label{sec:conclusion}

We present a learnable brain tumor model personalization methodology. We demonstrate that it is feasible to learn an inverse model in a supervised fashion from a data set of numerical simulations. We show that the choice of output can crucially affect network's performance - predicting time-independent combinations of parameters notably outperforms time-dependent counterparts. Such time-independent parametrization is not limited to the PDEs considered here, and thus the proposed \emph{Learn-Morph-Infer} pipeline can be adapted to other inverse problems in natural science and engineering disciplines. For the brain tumor growth model, the \emph{Learn-Morph-Infer} pipeline provides real-time performance of the parametric inference. Most importantly, the personalization time is stable across tumor models of different numerical complexity. These performance benefits pave the way for clinical testing of various mathematical tumor descriptions on a large cohort of patients.

\section{Acknowledgment}
I. Ezhov, S. Shit and L. Canalini are supported by the Translational Brain Imaging Training Network under the EU Marie Sklodowska-Curie program (Grant ID: 765148). B. Menze, B. Wiestler, and F. Kofler are supported through the SFB 824, subproject B12, by DFG through TUM International Graduate School of Science and Engineering, GSC 81, and by the Institute for Advanced Study, funded by the German Excellence Initiative. The authors acknowledge funding from DCoMEX (Grant agreement ID: 956201). B. Menze is supported by Helmut Horten Foundation.

\bibliography{mybib}

\begin{thebibliography}{46}
\providecommand{\natexlab}[1]{#1}
\providecommand{\url}[1]{\texttt{#1}}
\expandafter\ifx\csname urlstyle\endcsname\relax
  \providecommand{\doi}[1]{doi: #1}\else
  \providecommand{\doi}{doi: \begingroup \urlstyle{rm}\Url}\fi

\bibitem[Avants et~al.(2009)Avants, Tustison, and Song]{avants2009advanced}
B.~B. Avants, N.~Tustison, and G.~Song.
\newblock Advanced normalization tools (ants).
\newblock \emph{Insight j}, 2\penalty0 (365):\penalty0 1--35, 2009.

\bibitem[Dax et~al.(2021)Dax, Green, Gair, Macke, Buonanno, and
  Sch{\"o}lkopf]{dax2021real}
M.~Dax, S.~R. Green, J.~Gair, J.~H. Macke, A.~Buonanno, and B.~Sch{\"o}lkopf.
\newblock Real-time gravitational wave science with neural posterior
  estimation.
\newblock \emph{Physical review letters}, 127\penalty0 (24):\penalty0 241103,
  2021.

\bibitem[Ezhov et~al.(2020)Ezhov, Mot, Shit, Lipkov{\'a}, Paetzold, Kofler,
  Navarro, Pellegrini, Kollovieh, Metz, Wiestler, and
  Menze]{Ezhov2020GeometryawareNS}
I.~Ezhov, T.~Mot, S.~Shit, J.~Lipkov{\'a}, J.~C. Paetzold, F.~Kofler,
  F.~Navarro, C.~Pellegrini, M.~Kollovieh, M.~Metz, B.~Wiestler, and B.~Menze.
\newblock Geometry-aware neural solver for fast bayesian calibration of brain
  tumor models.
\newblock 2020.

\bibitem[Ezhov et~al.(2019)]{Ezhov_2019}
I.~Ezhov et~al.
\newblock Neural parameters estimation for brain tumor growth modeling.
\newblock In \emph{MICCAI}, pages 787--795. 2019.

\bibitem[Geremia et~al.(2012)Geremia, Menze, Prastawa, Weber, Criminisi, and
  Ayache]{geremia2012brain}
E.~Geremia, B.~H. Menze, M.~Prastawa, M.-A. Weber, A.~Criminisi, and N.~Ayache.
\newblock Brain tumor cell density estimation from multi-modal mr images based
  on a synthetic tumor growth model.
\newblock In \emph{International MICCAI Workshop on Medical Computer Vision},
  pages 273--282. Springer, 2012.

\bibitem[Harpold et~al.(2007)Harpold, Alvord~Jr, and
  Swanson]{harpold2007evolution}
H.~L. Harpold, E.~C. Alvord~Jr, and K.~R. Swanson.
\newblock The evolution of mathematical modeling of glioma proliferation and
  invasion.
\newblock \emph{Journal of Neuropathology \& Experimental Neurology},
  66\penalty0 (1):\penalty0 1--9, 2007.

\bibitem[He et~al.(2016)He, Zhang, Ren, and Sun]{deepreslearning}
K.~He, X.~Zhang, S.~Ren, and J.~Sun.
\newblock Deep residual learning for image recognition.
\newblock In \emph{2016 IEEE Conference on Computer Vision and Pattern
  Recognition (CVPR)}, pages 770--778, 2016.
\newblock \doi{10.1109/CVPR.2016.90}.

\bibitem[Hogea et~al.(2008)]{hogea_rd_mass}
C.~Hogea et~al.
\newblock An image-driven parameter estimation problem for a
  reaction--diffusion glioma growth model with mass effects.
\newblock \emph{J Math Biol}, 56\penalty0 (6):\penalty0 793--825, 2008.

\bibitem[Hormuth et~al.(2018)Hormuth, Eldridge, Weis, Miga, and
  Yankeelov]{hormuth2018mechanically}
D.~A. Hormuth, S.~L. Eldridge, J.~A. Weis, M.~I. Miga, and T.~E. Yankeelov.
\newblock Mechanically coupled reaction-diffusion model to predict glioma
  growth: methodological details.
\newblock In \emph{Cancer Systems Biology}, pages 225--241. Springer, 2018.

\bibitem[Hormuth et~al.(2019)Hormuth, Jarrett, Feng, and
  Yankeelov]{hormuth2019calibrating}
D.~A. Hormuth, A.~M. Jarrett, X.~Feng, and T.~E. Yankeelov.
\newblock Calibrating a predictive model of tumor growth and angiogenesis with
  quantitative mri.
\newblock \emph{Annals of biomedical engineering}, 47\penalty0 (7):\penalty0
  1539--1551, 2019.

\bibitem[Hormuth et~al.(2021)]{hormuth2021image}
D.~A. Hormuth et~al.
\newblock Image-based personalization of computational models for predicting
  response of high-grade glioma to chemoradiation.
\newblock \emph{Scientific Reports}, 11\penalty0 (1):\penalty0 1--14, 2021.

\bibitem[Jackson et~al.(2015)Jackson, Juliano, Hawkins-Daarud, Rockne, and
  Swanson]{Jackson_2015}
P.~R. Jackson, J.~Juliano, A.~Hawkins-Daarud, R.~C. Rockne, and K.~R. Swanson.
\newblock Patient-specific mathematical neuro-oncology: Using a simple
  proliferation and invasion tumor model to inform clinical practice.
\newblock \emph{BMB}, 77\penalty0 (5):\penalty0 846--856, mar 2015.

\bibitem[Kasim et~al.(2020)]{kasim2020up}
M.~Kasim et~al.
\newblock Up to two billion times acceleration of scientific simulations with
  deep neural architecture search.
\newblock \emph{arXiv preprint arXiv:2001.08055}, 2020.

\bibitem[Kim et~al.(2018)]{Kim_2019}
B.~Kim et~al.
\newblock Deep fluids: A generative network for parameterized fluid
  simulations.
\newblock \emph{Comput. Graph. Forum}, 38:\penalty0 59--70, 2018.

\bibitem[Kingma and Ba(2015)]{adam}
D.~P. Kingma and J.~Ba.
\newblock Adam: {A} method for stochastic optimization.
\newblock In \emph{ICLR 2015}, 2015.

\bibitem[Kofler et~al.(2020)]{kofler2020brats}
F.~Kofler et~al.
\newblock Brats toolkit: Translating brats brain tumor segmentation algorithms
  into clinical and scientific practice.
\newblock \emph{Frontiers in Neuroscience}, 14, 2020.

\bibitem[Konukoglu et~al.(2010{\natexlab{a}})]{KONUKOGLU2010111}
E.~Konukoglu et~al.
\newblock Extrapolating glioma invasion margin in brain magnetic resonance
  images: Suggesting new irradiation margins.
\newblock \emph{MedIA}, 14\penalty0 (2):\penalty0 111 -- 125,
  2010{\natexlab{a}}.

\bibitem[Konukoglu et~al.(2010{\natexlab{b}})]{ender_eikonal}
E.~Konukoglu et~al.
\newblock Image guided personalization of reaction-diffusion type tumor growth
  models using modified anisotropic eikonal equations.
\newblock \emph{IEEE Transactions on Medical Imaging}, 29\penalty0
  (1):\penalty0 77--95, 2010{\natexlab{b}}.

\bibitem[Le et~al.(2017)]{Le_tmi}
M.~Le et~al.
\newblock Personalized radiotherapy planning based on a computational tumor
  growth model.
\newblock \emph{{IEEE} Transactions on Medical Imaging}, 36\penalty0
  (3):\penalty0 815--825, mar 2017.

\bibitem[Lipkova et~al.(2019)]{jana_tmi}
J.~Lipkova et~al.
\newblock Personalized radiotherapy design for glioblastoma: Integrating
  mathematical tumor models, multimodal scans and bayesian inference.
\newblock \emph{IEEE Transactions on Medical Imaging}, pages 1--1, 2019.

\bibitem[Lorenzo and et~al(2021)]{lorenzo2021quantitative}
G.~Lorenzo and et~al.
\newblock Quantitative in vivo imaging to enable tumor forecasting and
  treatment optimization.
\newblock \emph{arXiv preprint arXiv:2102.12602}, 2021.

\bibitem[Loshchilov and Hutter(2019)]{loshchilov2018decoupled}
I.~Loshchilov and F.~Hutter.
\newblock Decoupled weight decay regularization.
\newblock In \emph{International Conference on Learning Representations}, 2019.
\newblock URL \url{https://openreview.net/forum?id=Bkg6RiCqY7}.

\bibitem[Lueckmann et~al.(2017)]{lueckmann_snpb}
J.-M. Lueckmann et~al.
\newblock Flexible statistical inference for mechanistic models of neural
  dynamics.
\newblock In \emph{NeurIPS}, pages 1289--1299, 2017.

\bibitem[Lê et~al.(2016)Lê, Delingette, Kalpathy-Cramer, Gerstner, Batchelor,
  Unkelbach, and Ayache]{matthieu_bayesian}
M.~Lê, H.~Delingette, J.~Kalpathy-Cramer, E.~R. Gerstner, T.~Batchelor,
  J.~Unkelbach, and N.~Ayache.
\newblock Mri based bayesian personalization of a tumor growth model.
\newblock \emph{IEEE Transactions on Medical Imaging}, 35\penalty0
  (10):\penalty0 2329--2339, 2016.
\newblock \doi{10.1109/TMI.2016.2561098}.

\bibitem[Mang et~al.(2012)Mang, Toma, Schuetz, Becker, Eckey, Mohr, Petersen,
  and Buzug]{mang2012biophysical}
A.~Mang, A.~Toma, T.~A. Schuetz, S.~Becker, T.~Eckey, C.~Mohr, D.~Petersen, and
  T.~M. Buzug.
\newblock Biophysical modeling of brain tumor progression: From unconditionally
  stable explicit time integration to an inverse problem with parabolic pde
  constraints for model calibration.
\newblock \emph{Medical Physics}, 39\penalty0 (7Part1):\penalty0 4444--4459,
  2012.

\bibitem[Menze et~al.(2011)]{bjoern_ipmi}
B.~Menze et~al.
\newblock A generative approach for image-based modeling of tumor growth.
\newblock In \emph{IPMI}, pages 735--747, 2011.
\newblock ISBN 978-3-642-22092-0.

\bibitem[Menze et~al.(2014)]{menzebrats}
B.~H. Menze et~al.
\newblock The multimodal brain tumor image segmentation benchmark (brats).
\newblock \emph{IEEE transactions on medical imaging}, 34\penalty0
  (10):\penalty0 1993--2024, 2014.

\bibitem[Papamakarios and Murray(2016)]{papamakarios_snpa}
G.~Papamakarios and I.~Murray.
\newblock Fast $\epsilon$-free inference of simulation models with bayesian
  conditional density estimation.
\newblock In \emph{NeurIPS}, pages 1028--1036. 2016.

\bibitem[Patel and Hathout(2017)]{patel2017image}
V.~Patel and L.~Hathout.
\newblock Image-driven modeling of the proliferation and necrosis of
  glioblastoma multiforme.
\newblock \emph{Theoretical Biology and Medical Modelling}, 14\penalty0
  (1):\penalty0 1--9, 2017.

\bibitem[Pati et~al.(2020)Pati, Sharma, Aslam, Thakur, Akbari, Mang,
  Subramanian, Biros, Davatzikos, and Bakas]{pati2020tmod}
S.~Pati, V.~Sharma, H.~Aslam, S.~Thakur, H.~Akbari, A.~Mang, S.~Subramanian,
  G.~Biros, C.~Davatzikos, and S.~Bakas.
\newblock Tmod-09. glioblastoma biophysical growth estimation using deep
  learning-based regression.
\newblock \emph{Neuro-Oncology}, 22\penalty0 (Supplement\_2):\penalty0
  ii229--ii229, 2020.

\bibitem[Raissi et~al.(2019)Raissi, Perdikaris, and
  Karniadakis]{raissi2019physics}
M.~Raissi, P.~Perdikaris, and G.~E. Karniadakis.
\newblock Physics-informed neural networks: A deep learning framework for
  solving forward and inverse problems involving nonlinear partial differential
  equations.
\newblock \emph{Journal of Computational Physics}, 378:\penalty0 686--707,
  2019.

\bibitem[Rohlfing et~al.(2010)Rohlfing, Zahr, Sullivan, and
  Pfefferbaum]{rohlfing2010sri24}
T.~Rohlfing, N.~M. Zahr, E.~V. Sullivan, and A.~Pfefferbaum.
\newblock The sri24 multichannel atlas of normal adult human brain structure.
\newblock \emph{Human brain mapping}, 31\penalty0 (5):\penalty0 798--819, 2010.

\bibitem[Scheufele et~al.(2019)Scheufele, Mang, Gholami, Davatzikos, Biros, and
  Mehl]{scheufele2019coupling}
K.~Scheufele, A.~Mang, A.~Gholami, C.~Davatzikos, G.~Biros, and M.~Mehl.
\newblock Coupling brain-tumor biophysical models and diffeomorphic image
  registration.
\newblock \emph{Computer methods in applied mechanics and engineering},
  347:\penalty0 533--567, 2019.

\bibitem[Scheufele et~al.(2020)Scheufele, Subramanian, and
  Biros]{scheufele2020automatic}
K.~Scheufele, S.~Subramanian, and G.~Biros.
\newblock Automatic mri-driven model calibration for advanced brain tumor
  progression analysis.
\newblock \emph{arXiv: Medical Physics}, 2020.

\bibitem[Sitzmann et~al.(2020)]{sitzmann2020implicit}
V.~Sitzmann et~al.
\newblock Implicit neural representations with periodic activation functions.
\newblock \emph{arXiv preprint arXiv:2006.09661}, 2020.

\bibitem[Stevens and Colonius(2020)]{stevens2020finitenet}
B.~Stevens and T.~Colonius.
\newblock Finitenet: A fully convolutional lstm network architecture for
  time-dependent partial differential equations.
\newblock \emph{arXiv preprint arXiv:2002.03014}, 2020.

\bibitem[Stupp et~al.(2014)Stupp, Brada, Van Den~Bent, Tonn, and
  Pentheroudakis]{stupp2014high}
R.~Stupp, M.~Brada, M.~Van Den~Bent, J.-C. Tonn, and G.~Pentheroudakis.
\newblock High-grade glioma: Esmo clinical practice guidelines for diagnosis,
  treatment and follow-up.
\newblock \emph{Annals of oncology}, 25:\penalty0 iii93--iii101, 2014.

\bibitem[Subramanian et~al.(2019)Subramanian, Gholami, and
  Biros]{subramanian2019simulation}
S.~Subramanian, A.~Gholami, and G.~Biros.
\newblock Simulation of glioblastoma growth using a 3d multispecies tumor model
  with mass effect.
\newblock \emph{Journal of mathematical biology}, 79\penalty0 (3):\penalty0
  941--967, 2019.

\bibitem[Subramanian et~al.(2020{\natexlab{a}})Subramanian, Scheufele,
  Himthani, and Biros]{subramanian2020multiatlas}
S.~Subramanian, K.~Scheufele, N.~Himthani, and G.~Biros.
\newblock Multiatlas calibration of biophysical brain tumor growth models with
  mass effect.
\newblock \emph{arXiv preprint arXiv:2006.09932}, 2020{\natexlab{a}}.

\bibitem[Subramanian et~al.(2020{\natexlab{b}})Subramanian, Scheufele, Mehl,
  and Biros]{subramanian2020did}
S.~Subramanian, K.~Scheufele, M.~Mehl, and G.~Biros.
\newblock Where did the tumor start? an inverse solver with sparse localization
  for tumor growth models.
\newblock \emph{Inverse Problems}, 36\penalty0 (4):\penalty0 045006,
  2020{\natexlab{b}}.

\bibitem[Swanson et~al.(2000)Swanson, Alvord~Jr, and
  Murray]{swanson2000quantitative}
K.~R. Swanson, E.~C. Alvord~Jr, and J.~Murray.
\newblock A quantitative model for differential motility of gliomas in grey and
  white matter.
\newblock \emph{Cell proliferation}, 33\penalty0 (5):\penalty0 317--329, 2000.

\bibitem[Thuerey et~al.(2020)Thuerey, Wei{\ss}enow, Prantl, and
  Hu]{thuerey2020deep}
N.~Thuerey, K.~Wei{\ss}enow, L.~Prantl, and X.~Hu.
\newblock Deep learning methods for reynolds-averaged navier--stokes
  simulations of airfoil flows.
\newblock \emph{AIAA Journal}, 58\penalty0 (1):\penalty0 15--26, 2020.

\bibitem[Tunc et~al.(2021)Tunc, Hormuth, Biros, and
  Yankeelov]{tunc2021modeling}
B.~Tunc, D.~Hormuth, G.~Biros, and T.~E. Yankeelov.
\newblock Modeling of glioma growth with mass effect by longitudinal magnetic
  resonance imaging.
\newblock \emph{IEEE Transactions on Biomedical Engineering}, 2021.

\bibitem[Tun{\c{c}} et~al.(2021)Tun{\c{c}}, Hormuth~II, Biros, and
  Yankeelov]{tuncc2021modeling}
B.~Tun{\c{c}}, D.~A. Hormuth~II, G.~Biros, and T.~E. Yankeelov.
\newblock Modeling of glioma growth with mass effect by longitudinal magnetic
  resonance imaging.
\newblock \emph{IEEE Transactions on Biomedical Engineering}, 68\penalty0
  (12):\penalty0 3713--3724, 2021.

\bibitem[Yankeelov et~al.(2013)Yankeelov, Atuegwu, Hormuth, Weis, Barnes, Miga,
  Rericha, and Quaranta]{yankeelov2013clinically}
T.~E. Yankeelov, N.~Atuegwu, D.~Hormuth, J.~A. Weis, S.~L. Barnes, M.~I. Miga,
  E.~C. Rericha, and V.~Quaranta.
\newblock Clinically relevant modeling of tumor growth and treatment response.
\newblock \emph{Science translational medicine}, 5\penalty0 (187):\penalty0
  187ps9--187ps9, 2013.

\bibitem[Young et~al.(1999)Young, Baum, Cremerius, Herholz, Hoekstra,
  Lammertsma, Pruim, Price, et~al.]{young1999measurement}
H.~Young, R.~Baum, U.~Cremerius, K.~Herholz, O.~Hoekstra, A.~Lammertsma,
  J.~Pruim, P.~Price, et~al.
\newblock Measurement of clinical and subclinical tumour response using
  [18f]-fluorodeoxyglucose and positron emission tomography: review and 1999
  eortc recommendations.
\newblock \emph{European journal of cancer}, 35\penalty0 (13):\penalty0
  1773--1782, 1999.

\end{thebibliography}

\end{document}